\documentclass[preprint,showpacs,showkeys,amsmath,amssymb]{revtex4}

\usepackage{graphicx}
\usepackage{dcolumn}
\usepackage{bm}

\begin{document}


\title{Relativistic Transport Approach to Collective Nuclear Dynamics}

\author{Serb\"{u}lent Yildirim }
\affiliation{
Fizik B\"{o}l\"{u}m\"{u}, Mersin \"{U}niversitesi, 33343 Mersin, Turkey
}
\author{Theodor Gaitanos}
\affiliation{
Physics Department, Universit\"at M\"unchen, D-85748 Garching, Germany
}
%
\author{Massimo Di Toro and Vincenzo Greco}
\email{ditoro@lns.infn.it}
\affiliation{
Laboratori Nazionali del Sud, INFN,  Via S. Sofia 62, I-95123 Catania, 
Italy\\
and Dipartimento di Fisica e Astronomia, Universita degli Studi di Catania}

\date{\today}

\begin{abstract}
The isoscalar giant monopole resonance $(ISGMR)$ and isovector giant dipole 
resonance $(IVGDR)$ in finite nuclei are studied in the framework of 
a relativistic transport approach. The kinetic equations are derived within
an effective nucleon-meson field theory in the Relativistic
Mean Field $(RMF)$ scheme, even extended to density dependent vertices. 
Small amplitude oscillations are analysed 
using the 
Relativistic Vlasov $(RV)$ approach, i.e. neglecting nucleon 
collision terms. 
The time evolution of the isoscalar 
monopole moment and isovector dipole moment and the corresponding Fourier 
power spectra are discussed. The excitation energies of $ISGMR$ and $IVGDR$
 are
 obtained for spherical nuclei with various sets of Lagrangian parameters.
 
In the case of $^{208}Pb$ we study in detail the dependence of the monopole
response on the effective mass and symmetry energy at saturation given
by the used covariant effective interaction. We show that a reduced
$m^{*}$ and a larger $a_4$ can compensate the effect on the $ISGMR$
energy centroid of a much larger compressibility modulus $K_{nm}$.
This result is important in order to overcome the conflicting
determination of the nuclear compressibility between non-relativistic
and relativistic effective interactions.

For the symmetry energy dynamical effects, we carefully analyze the 
influence of the inclusion of an 
effective isovector scalar channel,
 $\delta$-meson field, with 
constant and density dependent couplings. We show that the $\delta$-meson  
contribution, keeping fixed the equilibrium $a_4$ value, leads to a
 change in the isoscalar and isovector response of finite nuclei which 
results
in a reduction in the centroid energy of $ISGMR$ and $IVGDR$ for $^{208}$Pb.
All that in fact reveals the relevance of the $slope$ (or pressure) of the
symmetry energy at saturation on the $ISGMR$ and $IVGDR$ modes for 
neutron-rich systems.

Density dependent vertices are not much affecting our conclusions. 
Following as a guidance some extended dispersion relations in nuclear
matter, we see two 
main reasons for that, the smoothness of the density dependences around
saturation and the presence of compensation effects coming from 
rearrangement terms.
\end{abstract}

\maketitle

\section{\label{sec:level1}INTRODUCTION}

In this work we study the isoscalar giant monopole and isovector giant dipole
 oscillations in $^{208}Pb$ using the Relativistic Vlasov method $(RV)$
in a Relativistic Mean Field $(RMF)$ theory with constant and 
density dependent 
meson-nucleon vertices. We analyze various $RMF$ models with different 
coupling
 parameters which lead to a different nucleon effective mass ($m^{*}$), 
compressibility modulus ($K_{nm}$), symmetry energy ($a_{4}$),  
keeping fixed other nuclear matter saturation properties. 

Among 
those   
models we focus on the ones where the  scalar isovector channel has been 
introduced by 
the coupling to an effective $\delta$-meson field  \cite{Liu_1}.
It has been shown that such inclusion  has important
 effects on the equation of state $(EOS)$ and phase diagram of asymmetric 
nuclear matter $(ANM)$ \cite{Liu_1}, as well as on the reaction dynamics
with exotic nuclei, see the review \cite{BaranPR410}. In fact the 
$\delta$-meson brings contributions to
 the slope and curvature of the symmetry energy and to the neutron-proton 
effective mass splitting. In particular the influence of this coupling on the
 collective response of $(ANM)$ appears important, as shown in a linear 
response approach in ref. \cite{Greco_1}. 
 For the comparison of the 
results we have also considered 
the well known $NL3$ \cite{Ring,Lalazissis} parametrization, very successful
 for finite nuclei structure calculations,  and some density 
dependent vertex models \cite{FuchsPRC52,Typel} beyond $RMF$. In this way 
we aim to 
pin down the sensitivity 
of  isoscalar and isovector collective oscillations on $m^{*}$, $K_{nm}$ 
and $a_{4}$, focussing on the interplay between isoscalar-isovector couplings.
In particular we will see that the covariant structure of the isovector 
part of the effective interaction will clearly show up even in the
isoscalar monopole mode in $^{208}Pb$. 

Relativistic mean field models, with constant \cite{Stoitsov_1,Vretenar_1} 
and density dependent \cite{Vretenar_2} meson-nucleon couplings, have
 been  applied to the
 description of collective excitations of atomic nuclei in the framework of 
Time-Dependent Relativistic Mean Field $(TDRMF)$ and in the self-consistent 
relativistic random phase approximation $(RRPA)$.
 The monopole predictions of nonrelativistic Hartree-Fock plus
 random phase approximation $(RPA)$ calculations, using both Skyrme and Gogny 
effective interactions, seem to indicate that the value of $K_{nm}$ should 
be in 
the range 210-220 MeV \cite{Blaizot_1,Blaizot_2}. In relativistic 
mean-field models on the other hand, results of both time-dependent and
 $RPA$ calculations suggest that empirical $GMR$ energies are best reproduced 
by an effective force with $K_{nm}$ 250-270 MeV 
\cite{Vretenar_4,Ma}. This difference has been pointed 
out as a serious open problem \cite{Vretenar_2}. A possible solution to this 
ambiguity was suggested in Ref.\cite{Greco_1} as an interplay between the 
effective mass and compressibility within a discussion of isoscalar response
 of nuclear matter. Here we show the validity of this interpretation,
 joint to a density dependence of the symmetry energy , also  
for the spherical finite n-rich nuclei, like $^{208}Pb$. Recent studies are 
in fact pointing to the same isospin effect, see refs. 
\cite{PiekaPRC66,AgraPRC68,colo}. In this respect we have even analyzed
the influence of the
 inclusion of the $\delta$-meson on both isoscalar and isovector response in 
spherical n-rich nuclei.

The introduction of density dependent vertices, following the $Dirac-Brueckner-
Hatree-Fock$ estimations, is not showing relevant new effects. A nice 
analitical interpretation of this result, based on a linear response theory
for nuclear matter, is finally presented.

\section{\label{sec:level2}OUTLINE OF THE MODEL}

The dynamics of collective vibrations in spherical nuclei is studied 
in the 
framework of the relativistic Vlasov ($RV$) transport equation, which 
describes the dynamical 
evolution of a semi-classical phase space distribution function $f(x,p)$ 
under 
the influence of the nuclear mean field. Binary collisions between nucleons 
will be not considered here. Thorough derivations of the $RV$ 
transport equations from an effective hadron-meson field theory, \cite{qhd} 
can be 
found elsewhere \cite{Botermans_1,Blaettel,Danielewicz}. The $RV$ equation 
reads 
($i=p,n$):
\begin{equation} \label{rv}
[p_i^{*\mu}\partial_{\mu}+(p^{*}_{\nu i}
\mathcal{F}_i^{\mu\nu}+m_i^{*}\partial^{\mu}m_i^{*})
\partial_{\mu}^{(p_{*})}]f_i(x,p^{*})=0
\end{equation}
whith the field tensor 
\begin{equation}
\mathcal{F}_i^{\mu\nu}\equiv\partial^{\mu}\Sigma_i^{\nu}-\partial^{\nu}
\Sigma_i^{\mu}
\end {equation}
and effective masses and kinetic momenta $m^{*}$ and $p^{*\mu}$, respectively,
 specified 
below. The particles obey the mass-shell condition 
\begin{equation} \label{mshell}
p_i^{*\mu}p^{*}_{i \mu}-m_i^{* 2}=0.
\end{equation}

Thus, from the temporal knowledge of the phase space distribution function one 
can calculate the time evolution of physical quantities such as densities and 
fields. We remind the meaning of the 
the Wigner matrix (in the Spinor space)  
\begin{widetext}
\begin{eqnarray} \label{eqn:1}
F_{\alpha\beta}(x,p)&=&<\hat{F}_{\alpha\beta}(x,p)> \nonumber \\
                    &\equiv& \frac{1}{(2\pi)^4}\int d^4R \mbox{ }e^{-ip_{\mu}
R^{\mu}} 
<\bar{\psi}_{\beta}(x+\frac{1}{2}R)\psi_{\alpha}(x-\frac{1}{2}R)>. 
\end{eqnarray}
\end{widetext}
>From the above 
definition it follows that single-particle operators can be 
expressed as (spinor indices will be omitted for simplicity) 
\begin{equation}
<\hat{O}>=\int d^4x \int d^4p \mbox{ tr}\Big(\hat{O}F(x,p)\Big)
\end{equation}
where the trace runs over spin and isospin indices. 
The scalar density and the vector current, for example, assume the form 
{\setlength\arraycolsep{2pt}
\begin{eqnarray} 
\rho_s(x)&=&<\bar{\psi}\psi>=\int d^4p \mbox{ tr}(F(x,p)) 
\label{eqn:sd} \\
j_{\mu}(x)&=&<\bar{\psi}\gamma_{\mu}\psi>=\int d^4p \mbox{ tr}(\gamma_{\mu}
F(x,p)), 
\label{eqn:cb}
\end{eqnarray}}
\noindent
and will be used to calculate the different Lorentz components of the mean 
field 
potential.

The nuclear mean field $U$ is characterized in a $RMF$ theory by means of self 
energies 
in the form $U=\Sigma_{s}-\gamma_{\mu} \Sigma^{\mu} \cdots$ (higher 
contributions are usualy neglected due to symmetry properties of nuclear 
matter). In a Non-Linear  ($QHD-NL$) model with isoscalar scalar and vector 
meson fields $\sigma$ and $\omega$ and with the inclusion of the isovector
 channel through the exchange of the virtual charged $\delta$(scalar) and 
$\rho$(vector) 
mesons, the mean field approximation leads to self energies which are related
 to the expectation values of the combination of isoscalar and isovector
 fields with coupling constants $g_{\sigma}$, $g_{\omega}$, $g_{\rho}$ and
 $g_{\delta}$. The scalar and vector components of the self energies are 
generally given by

\begin{equation}
\label{self-energy}
\Sigma_i^{\mu}= g_{\omega}\omega^{\mu}(x)\pm g_{\rho}b^{\mu}(x) \qquad
\left\{
\begin{array}{rl}
+ & \mbox{proton} (i=p)\\
- & \mbox{neutron } (i=n)
\end{array}
\right.
\end{equation}

\begin{equation}
\Sigma_{si}= g_{\sigma}\sigma(x)\pm g_{\delta}\delta(x) \qquad
\left\{
\begin{array}{rl}
+ & \mbox{proton } (i=p)\\
- & \mbox{neutron } (i=n)
\end{array}
\right.
\end{equation}
\noindent
with the expectation values of the fields self-consistently calculated,
 see later Eqs.(\ref{fields}).

The self energies characterize the in-medium properties of the 
nucleons inside the hadronic environment in terms of kinetic 
momenta and effective masses
\begin{equation}
\label{peff}
p_i^{*\mu}= p_i^{\mu}-\Sigma_i^{\mu}
\end{equation}
\begin{equation}
m_i^{*}= m-\Sigma_{si}
\end{equation}

The density dependence of the mean field, i.e. the density behavior of 
the self energies, depends on the coupling choices of the $RMF$ model. Here 
we consider 
different parametrizations within the Non-Linear ($QHD-NL$) effective 
field approach
 and even extended to the Density Dependent Hadronic ($DDH$) mean field 
theory \cite{FuchsPRC52,Typel,Gaitanos}. 
In table \ref{table1} the values for the different coupling 
constants and the non-linear parameters for different sets of Non-Linear 
Walecka ($QHD-NL$) models are presented, for details see refs.
 \cite{Liu_1,BaranPR410,Greco_1,Ring,Lalazissis,Typel}. 
Their corresponding nuclear matter saturation 
properties are given in table \ref{table2}. In the $DDH$ models the density
dependence of the coupling constants is chosen in order to reproduce
microscopic $Dirac-Brueckner-Hartree-Fock$ results beyond the $RMF$ picture,
 see the discussion in ref. \cite{Gaitanos}.

The choice of models with rather 
different nuclear matter properties has been done on purpose, in order to 
investigate 
the role of the effective masses, compressibility and symmetry energy on
isoscalar (monopole) and isovector (dipole) oscillations in neutron-rich
nuclei.
We have compared various Non-Linear $RMF$ parametrizations, in particular the 
Giessen sets $NL1-G, NL2-G$, \cite{Maruyama}, extended also to simulations of
relativistic heavy ion collisions, and the widely used $NL3$ set 
\cite{Ring,Lalazissis}, successfully applied in 
finite nuclei studies. The different treatment of the iso-vector part of the 
mean field 
(competition effects of the repulsive $\rho$ field and the attractive $\delta$
 meson) is analysed in detail using 
the ($NL\rho,~NL\rho\delta$) parameter 
sets \cite{Liu_1}. The same analysis is performed within density dependent
coupling models, the parametrizations $DDH3\rho$ and $DDH3\rho\delta$
of ref.\cite{Gaitanos} have been used.

\begin{table}[t]
\caption{\label{table1}Coupling parameters in terms of 
$f_{i}\equiv(g_{i}/m_{i})^{2} $ 
for $i=\sigma, \omega$, $f_{i}\equiv(g_{i}/2m_{i})^{2} $
for $\rho, \delta$, $A \equiv a/g^{3}_{\sigma}$ and 
$B \equiv b/g^{4}_{\sigma}$ for various Non-Linear $RMF$ models using 
the $\rho$ 
and both $\rho$ and $\delta$ mesons for the characterization of the isovector 
mean field. In the $DDH$ models the coupling 
functions 
are explicitly density dependent \cite{FuchsPRC52,Typel,Gaitanos}.}
\begin{ruledtabular}
\begin{tabular}{cccccc}
 Parameter&NL1-G&NL2-G&NL3&
 $NL\rho$&$NL\rho\delta$\\
\hline
$f_{\sigma}$($fm^{2}$)& 6.146 & 9.300 & 15.739 &10.330
& 10.330  \\
$f_{\omega}$($fm^{2}$)& 3.611 & 3.611 & 10.530 &5.420
&5.420 \\
$f_{\rho}$($fm^{2}$)& 1.200 & 1.220 & 1.339 &0.950
& 3.150 \\
$f_{\delta}$($fm^{2}$)& 0.0 & 0.0 & 0.0 &0.0
& 2.500 \\
$A$($fm^{-1}$)& -0.123 &0.0824 &-0.01 &0.033 
& 0.033 \\
$B$& 0.169 &-0.0012 &-0.003 &-0.0048
& -0.0048  \\
\end{tabular}
\end{ruledtabular}
\end{table}
\begin{table}[t]
\caption{\label{table2}Nuclear matter saturation properties in the different 
$RMF$ models.}
\begin{ruledtabular}
\begin{tabular}{cccccc}
 Property & NL1-G&NL2-G&NL3&
 $NL\rho,\rho\delta$ & $DDH\rho,\rho\delta$\\
\hline
$E/A(MeV)$ &-16.0&-16.0&-16.3&-16.0&-16.0 \\
$\rho_{0}$$(fm^{-3})$ &0.145 &0.145 & 0.148 &0.160
& 0.153  \\
$m^{*}/m$& 0.83 & 0.83 & 0.60 &0.75
&0.55 \\
$K_{nm}$$(MeV)$& 380 & 210 & 272 &240
& 240 \\
$a_{4}$($MeV$)& 30.62 & 30.62 & 37.40 &30.50
& 33.40 \\
\end{tabular}
\end{ruledtabular}
\end{table}

The dynamics of collective vibrations in spherical nuclei is simulated 
in the framework of the relativistic Vlasov equation (\ref{rv}). 
Its numerical realization is based on the standard test particle method, 
where the 
phase space distribution function $f(x,p)$ is represented by a finite 
number of test particles of a covariant Gaussian form (Relativistic 
Landau-Vlasov ($RLV$) method). A detailed description of the $RLV$ method is 
given in Ref. \cite{Fuchs_1}, here this model will be briefly discussed. 
The use of a Gaussian shape for the test particles has the advantage of 
smooth distribution functions, but maintaining an accurate determination of 
local 
quantities, particularly important near the nuclear surface. 

The covariant Gaussians 
in the four-dimensional Minkowski space are defined as
\begin{widetext}
\begin{eqnarray}\label{gaussx}
G(x;\xi_i)&\equiv & \int_{-\infty}^{-\infty}d\tau g(x-x_i(\tau)) \nonumber\\
&=& \alpha\int_{-\infty}^{-\infty}d\tau \mbox{ exp}((x-x_i(\tau))^2/w^2)
\delta[(x_{\mu}-x_{i\mu}(\tau))u_i^{\mu}(\tau)]
\end{eqnarray}
\end{widetext}
where $\xi_{i}$ denotes the world line of the particle $i$ as a whole, $\tau$ 
refers to the eigentime of the test particle and $\alpha$ is the normalization 
constant. In the four-dimensional momentum space a gaussian weight of a test 
particle is defined by \cite{Fuchs_1}
\begin{equation}\label{gaussp}
g(p^{*}-p_i^{*}(\tau))\equiv\alpha_p 
\mbox{ exp}((p^{*}-p_i^{*}(\tau))^2/w_p^2)
\delta[p_{\mu}^{*}p_i^{*\mu}(\tau)-m_i^{*2}]
\end{equation}
where the center of the gaussian is assumed to be on-shell  i.e.  
$p_{i\mu}^{*}=m_i^{*}u_{i\mu}$, $u_i^2=1$, whereas the free momentum 
$p_{\mu}^{*}$ is generally off-shell. The effective mass of the particle 
is 
taken as $m_i^{*}=m^{*}(x_i(\tau))$. The norm of the gaussian is 
calculated in the rest frame of the particle to be $m^{* -1}_{i}$ 
with $\alpha_{p}=(\sqrt{\pi}w_{p})^{-3}$, where $w$ and $w_{p}$ 
are the test particle widths in position and momentum space. 

With the Gaussians of (\ref{gaussx},\ref{gaussp}) the phase space distribution 
function $f(x,p^{*})$ is expressed as
\begin{eqnarray} \label{distr}
f(x,p^{*})&=&\frac{1}{N}\sum_{i=1}^{A \cdot N} 
\int_{-\infty}^{+\infty}d\tau g(x-x_i(\tau))g(p^{*}-p_i^{*}(\tau))  
       \nonumber \\
              &=&\frac{1}{N(\pi w w_p)^3}
\sum_{i=1}^{A\cdot N} \int_{-\infty}^{-\infty}d\tau\mbox{ exp}((x-x_i
(\tau))^2/w^2) 
     \nonumber \\
              &{}&\times \mbox{ exp}((p^{*}-p_i^{*}(\tau))^2
/w_p^2)    
\nonumber \\
              &{}&\times \mbox{ }\delta[(x_{\mu}-x_{i\mu}(\tau))u_i^{\mu}
(\tau)] 
\delta[p_{\mu}^{*}p_i^{*\mu}(\tau)-m_i^{* 2}]
\end{eqnarray} 
where $N$ is the number of test particles per nucleon. Scalar densities
 $\rho_{s}$ 
and baryon currents $j_{\mu}$ follow from the phase space distribution  
\begin{eqnarray}
\rho_s(x)&=&\frac{1}{N}\sum_{i=1}^{A\cdot N}\int_{-\infty}^{-\infty}d\tau 
\frac{m^{*}(x)}{m^{*}(x_i(\tau))}g(x-x_i(\tau))  \label{dens} \\
j_{\mu}(x)&=&\frac{1}{N}\sum_{i=1}^{A\cdot N}
\int_{-\infty}^{-\infty}d\tau g(x-x_i(\tau))u_{i\mu}(\tau). \label{jmu2}
\end{eqnarray}

The equations of motion for the test particle trajectories are given by
\begin{widetext}
\begin{eqnarray}
\frac{d}{d\tau}x_i^{\mu}(\tau)&=&u_i^{\mu}(\tau)  \\
\frac{d}{d\tau}u_i^{\mu}(\tau)&=&\frac{1}{m^{*}(x_i)}\sum_{j=1}^{A\cdot N}
\frac{2}{w^2}
\Big[\frac{g_{\omega}^2}{m_{\omega}^2}u_{i\nu}(R_j^{\mu}(x_i)u_j^{\nu}-
R_j^{\nu}(x_i)u_j^{\mu}) 
\nonumber  \\
      &{}&
-g_{\sigma}\frac{\partial \sigma(x_i)}{\partial \rho_s}(R_j^{\mu}(x_i)-u_i^
{\mu}u_i^{\nu}R_{j\nu}(x_i)) 
\Big]\frac{\mbox{exp}(R_j^2(x_i)/w^2)}{N(\sqrt{\pi}w)^3} \nonumber \\
      &\pm& \frac{1}{m^{*}(x_i)}\frac{2}{w^2}\sum_{j=1}^{Z\cdot N}
\Big[\frac{g_{\rho}^2}{4m_{\rho}^2}u_{i\nu}(R_j^{\mu}(x_i)u_j^{\nu}-R_j^{\nu}
(x_i)u_j^{\mu})
 \nonumber \\
      &{}&
-\frac{g_{\delta}^2}{4m_{\delta}^2}u_{i\nu}(R_j^{\mu}(x_i)-u_i^{\mu}u_i^{\nu}
R_{j\nu}(x_i)) 
\Big] \frac{\mbox{exp}(R_j^2(x_i)/w^2)}{N(\sqrt{\pi}w)^3} \nonumber
 \\
      &\mp& \frac{1}{m^{*}(x_i)}\frac{2}{w^2}\sum_{j=Z\cdot N+1}^
{A\cdot N}
\Big[\frac{g_{\rho}^2}{4m_{\rho}^2}u_{i\nu}(R_j^{\mu}(x_i)u_j^{\nu}-R_j^{\nu}
(x_i)u_j^{\mu})
 \nonumber \\
      &{}&
-\frac{g_{\delta}^2}{4m_{\delta}^2}u_{i\nu}(R_j^{\mu}(x_i)-u_i^{\mu}u_i^{\nu}
R_{j\nu}(x_i))
\Big]
\frac{\mbox{exp}(R_j^2(x_i)/w^2)}{N(\sqrt{\pi}w)^3} \label{rlv}
\end{eqnarray}
\end{widetext}
with
$$
R_i^{\mu}(x) \equiv (x^\mu-x_i^\mu(\tau)) - 
(x_\nu -x_{i\nu}(\tau))u_i^\nu(\tau)u_i^\mu(\tau),
$$
projection of the vector $(x-x_i(\tau))$ on the hyperplane perpendicular
to $u_i(\tau)$, \cite{Fuchs_1}.

Here the equations for velocities, rather than for momenta, are 
given, within the 
assumption that the particle accelerations are small. In this equation  
particles 
are propagated by their respective eigentimes and so corresponding time 
coordinates $x_{i0}(\tau)$ 
 can differ. In order to solve the problem of different time 
coordinates 
a $system~time$ has been adopted for the propagation \cite{Fuchs_1}.

The scalar and vector meson fields, determined by the scalar density 
$\rho_{s}$ and the baryonic current $j_{\mu}$, respectively, result 
from the solution of the corresponding equations in the Local Density 
Approximation ($LDA$) 
\begin{widetext}
\begin{eqnarray}
m_\sigma^2\sigma(x)+B\sigma^2(x)+C\sigma^3(x)&=&g_\sigma\rho_s(x) 
 \equiv g_{\sigma} 
\int d^4p^{*} \frac{m^{*}(x)}{E^{*}(x)} f(x,p^{*})      
  \nonumber \\
\omega_{\mu}(x)&=& \frac{g_{\omega}}{m_{\omega}^2}j_{\mu}(x) \equiv
\frac{g_{\omega}}{m_{\omega}^2} \int d^4p^{*}p_{\mu}^{*}
f(x,p^{*})     
 \nonumber \\
b_{\mu}(x)&=& \frac{g_{\rho}}{4m_{\rho}^2}j_{3 \mu}(x) \equiv
\frac{g_{\rho}}{4m_{\rho}^2}\int d^4p^{*}p_{\mu}^{*} f_3(x,p^{*})     
 \nonumber \\
\delta(x)&=& \frac{g_{\delta}}{4m_{\delta}^2} \rho_{s3}(x)
\equiv  \frac{g_{\delta}}{4m_{\delta}^2}
\int d^4p^{*} \frac{m^{*}(x)}{E^{*}(x)} f_3(x,p^{*}). 
\label{fields}
\end{eqnarray}
\end{widetext}
with $f_3 \equiv f_p - f_n$.

An important issue for the description of low energy excitations within the 
$RV$ equation is an appropriate initialization of finite nuclei, before 
starting the phase space evolution of the distribution function with 
Eq. (\ref{rlv}). The ground state of a spherical nucleus is randomly 
initialized 
by means of the test particles with the covariant Gaussian shape in position
 and 
momentum space. After the first step of a randomly obtained initial 
distribution, 
a fit procedure in coordinate space is performed according to realistic 
Thomas-Fermi reference density distributions. During this procedure the 
proton and neutron distributions are fitted separately by satisfying the 
empirical 
values for the asymmetry and the surface thickness parameters, respectively. 
In the Thomas-Fermi calculation the scalar densities are determined by solving 
self-consistently the equations for the effective masses $m^{*}_{p,n}$. 
With this initialization of a given spherical nucleus the temporal 
evolution is described  
by the $RLV$ equation (\ref{rlv}). We have used $100$ test particles per 
nucleon for 
the transport descriptions, which yields a smooth distribution function with a 
very good energy conservation.

For giant resonances the time dependence of collective dynamical quantities 
is not 
periodic, since giant resonances are generally not stationary states of the 
mean-field Hamiltonian \cite{Vretenar_1}. For non-relativistic models it has 
been 
proven that the results from the Vlasov equation are identical to results from 
full quantum mechanical calculations ($TDHF$) \cite{Stocker_1}. Since in the 
non-relativistic frame $RPA$ is the small amplitude limit of a $TDHF$ 
calculation, 
one can expect that the results from the relativistic Vlasov equation are 
comparable 
to those from relativistic $RPA$ calculations of the type discussed in 
Ref. \cite{Vretenar_4}. In the small amplitude limit the energy 
obtained from the frequency of the oscillation coincides with the experimental 
energy of the collective oscillation.

The collective dynamical variables that characterize nuclear vibrations 
are defined as expectation values of single particle operators in the phase 
space representation. For the isoscalar monopole vibrations, the time 
dependent monopole 
moment is defined as 
\begin{equation}\label{isos}
\left\langle r^{2}(t)\right\rangle=\frac{1}{A}\int d^{3}x r^2 j_{0}(x). 
\end{equation} 

In this work we have applied the $RLV$ method to isoscalar monopole and 
isovector 
dipole oscillations in $^{208}Pb$. The excitation of an iso-scalar monopole 
oscillation of the initialized nucleus in its ground state is modeled by a 
radial expansion. This is done by introducing a new coordinate 
\begin{equation}
r_{mon}=(1+a)r
\end{equation}
for each test particle. Here $a$ is a scaling parameter, $a=0.1fm$ has been 
used, and the deformation of protons and neutrons is in phase. 

For the isovector dipole oscillation the following operator has been applied 
\cite{Vretenar_7}
\begin{equation} \label{isov}
\hat{Q}^{T=1}_{1\mu}=  \frac{N}{N+Z}\sum_{p = 1}^{z}r_{p}{Y_{1\mu }}
-\frac{Z}{N+Z}\sum_{n = 1}^{N} r_{n}Y_{1\mu }.
\end{equation}      
This means an out of phase shift along the z-axis between protons and 
neutrons. We have used a 
scaling parameter of $1~fm$ according to Eq.(\ref{isov}), which causes a 
center of mass separation between protons and neutrons, while keeping 
unchanged the center of mass of the whole system.

\section{\label{sec:level3}ISOSCALAR MONOPOLE OSCILLATIONS}

\begin{figure*}
\includegraphics[scale=0.6,angle=-90]{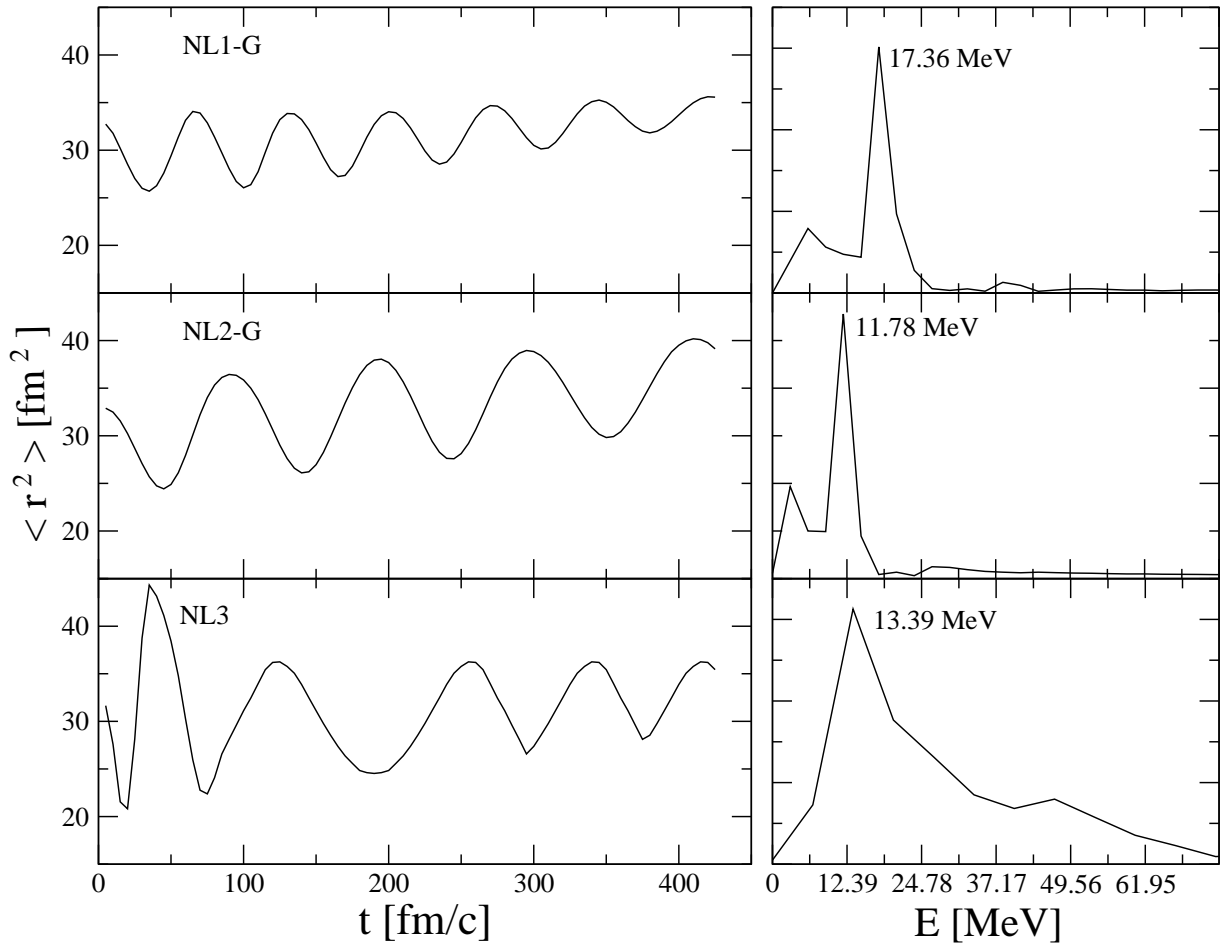}
\caption{\label{fig:1} (Left) Time-dependent isoscalar 
$\left\langle r^{2}\right\rangle$ 
monopole moment and (right) the corresponding Fourier power spectra for 
$^{208}$Pb, as 
obtained from transport calculations with different choices of the 
nuclear mean field: 
$(NL1-G, NL2-G, NL3)$ non-linear parametrizations of Refs. 
\cite{Maruyama,Ring}, respectively. 
The corresponding excitation energies of the $ISGMR$ are indicated in the right
 pannels.}
\end{figure*}

The study of Isoscalar Giant Monopole Resonances, $ISGMR$, in nuclei 
provides an important source of 
information on the nuclear matter compressibility. The complete experimental 
data set on isoscalar monopole 
 resonances 
has been analyzed by Shlomo and Youngblood \cite{Shlomo}. In fact, within a 
semi-empirical macroscopic 
approach of the systematics of $GMR$, it turns out that starting from finite
 nuclei data the nuclear matter 
compression modulus 
$K_{nm}$ can be only fixed within the rather wide range $200-350$ MeV. 
A systematic theoretical work, based on non-relativistic $RPA$ methods
with realistic effective interactions \cite{Blaizot_2}, was supporting
the lower region of the above range, around $210-220~MeV$. At variance
relativistic approaches, within the same small amplitude limits, were
pointing to some definitely larger values, $250-270~MeV$ \cite{Vretenar_2}.
We address the problem studying in a relativistic frame the $ISGMR$ in 
$^{208}Pb$, that shows a 
well established Giant Monopole Resonance ($GMR$) at  $13.7 \pm 0.3$ MeV.
Moreover  $^{208}Pb$ is interesting since it represents a neutron-rich
nucleus, where we can even reveal isovector effects.

As discussed in the previous section, the   
iso-scalar monopole oscillations are analyzed within the relativistic 
Vlasov transport equation. 
The nuclear mean field is evaluated in the Non-Linear versions of the 
Walecka model and 
in the $DDH$ approach, see Tables \ref{table1} and \ref{table2}. We will 
discuss the dependence 
of the $GMR$ on the compression modulus, e.g. by comparing the Giessen 
parametrizations 
(NL1-G and NL2-G) with a similar value for $m^{*}$ but different $K_{nm}$,
 and on the effective mass (for fixed $K_{nm}$), e.g. by comparing the 
parametrizations 
of Liu et al. with the $DDH$ model. The latter will allow also some comments 
on the effect of a density dependence of the couplings around the 
saturation point.
These analyses are comparable with similar 
studies in the 
framework of a time dependent relativistic mean field model, see Refs. 
\cite{Ring,Podobnik,Vretenar_4}, and within non-relativistic 
RPA calculations of the nuclear matter compressibility using Gogny effective 
interactions, see 
Ref. \cite{Blaizot_2}. 

The influence of a different treatment of the 
iso-vector part of the 
mean field will be particularly discussed either paying attention to the
$a_4$ differences of the various interactions, see Table \ref{table2},
 or more specifically focussing on the effects of the iso-vector, scalar 
$\delta$ field, 
i.e. by comparing the results of parametrizations, like 
$NL\rho,NL\rho\delta$ and $DDH\rho,DDH\rho\delta$, with  
and without the introduction of the $\delta$ meson. 
We note again that 
the inclusion of the $\delta$ meson in the interaction leads to a stiffer 
symmetry energy (around saturation) and to a splitting between the 
(Dirac) effective masses 
of neutrons and 
protons, see a review in ref. \cite{BaranPR410}. Also the vector component of 
the iso-vector self energy is modified 
due to the 
enhancement of the vector, iso-vector $\rho$ meson field roughly by a 
factor of $3$. 

Fig. 1 shows the dynamical evolution of the monopole moment of the excited 
${}^{208}Pb$ nucleus 
for those models with fixed effective mass $(NL1-G, NL2-G)$ and different 
compression 
modulus. The $ISGMR$ oscillation is modeled by Eq. (\ref{isos}). The numerical
 simulation of 
the $RV$ equation is performed with the test particle method of the previous 
section. The limited 
number of test particles per nucleon ($100$ in our case) leads to numerical 
fluctuations, 
which can be seen in Fig. \ref{fig:1} in terms of spurious oscillations with
 very low frequencies (first peaks in the Fourier energy spectra) and in 
terms of a 
partially non-periodic evolution  of the monopole moment. This is a general 
non-trivial 
feature of transport descriptions with test particle methods which leads to 
an artificial 
damping of the excitation \cite{CT}. However, due to a rather good stability 
of the 
nucleus (until several $100$ fm/c) and a good separation of the ambigous 
excitation energy 
of the numerical oscillations, we conclude on the reliability of the results 
presented 
in the following. 

The excitation energy of 
the ISGMR oscillation is very sensitive to the nuclear matter compressibility,
 which is 
a well known fact. In particular, the excitation energy increases with rising 
compression 
modulus (from $NL2-G$ to $NL1-G)$), as expected. 

The experimental  value of $13.7 \pm 0.3 MeV$ can be 
reproduced 
satisfactory by the $NL3$ model 
($13.39~(\hbar \omega)$). This result is consistent with the 
analysis of Ref. \cite{Ring}, again as an important check of the transport 
calculations.

We notice however that
the $NL2-G$ model, with $K_{nm}=210$ MeV, leads to a centroid  
excitation energy 
($11.78~(\hbar \omega)$) not too far from that of the $NL3$ model,
which has a much larger compressibility, $K_{nm}=272$ MeV.
Considering the other saturation properties of the $NL3$ force, see
Table \ref{table2}, this fact is pointing to two interesting implications:
i) The dependence of the monopole frequency on the effective nucleon mass;
ii) The dependence on the symmetry energy, the $a_4$ parameter, that one 
could expect since the $^{208}Pb$ is a neutron-rich system. We will discuss
in detail the two effects in the following. We note that, as a straightforward
consequence, the claimed discrepancy on the nuclear matter compressibility 
modulus $K_{nm}$
 between relativistic and non-relativistic models, see \cite{Vretenar_2}, 
can be eventually overcome.

\subsubsection{Monopole frequency and effective mass}
 
A linear response theory for nuclear matter within the $RMF$ frame can give
some interesting hints.
We can derive a dispersion relation, see refs. \cite{Greco_1,BaranPR410},

\begin{equation}\label{mondis}
1+ \frac{E^*_F}{3\,k_F^2}\left[K_{nm}^{pot}-9\,f_{\omega}
\frac{k^2_F}{E^{*2}_F}\left(1-f_\sigma\,\frac{m^*}{E^{*2}_F}\,\rho_S\right)
\rho_B\right]\varphi(s)=0\,.
\end{equation}
where $s$ is the dispersion parameter
$s \equiv \frac{v_s}{v_F} = \frac{\omega}{k \cdot v_F} $ and $\varphi(s)$ is 
the usual 
Lindhard function of the
Landau Fermi Liquid  theory:
$$\varphi(s)=1-{\frac{s}{2}}ln\left|{\frac{s+1}{s-1}}\right|+{\frac{i}{2}}
\,\pi s\,\theta(1-s)~$$
Here the $K_{nm}^{pot}$ is the potential part of the nuclear matter 
compressibility.

From Eq.(\ref{mondis}) we see that the ``restoring'' force for monopole
oscillations is given by an effective compressibility which is reduced for 
larger values of the $\omega$ meson  coupling constant.
 However $f_\omega$ can assume
very different values depending on the chosen value for effective masses
$m^*$. 
This is easy to understand since in the $RMF$ limit the saturation binding
energy has the simple form 
$$E/A(\rho_0)=E^*_F+f_\omega\rho_0-m_N$$
where $m_N$ is the bare nucleon mass. Thus in order to
have the same saturation values
of $\rho_0$, $E/A(\rho_0)$, when we decrease $m^*$ we have to 
increase $f_\omega$.
We then come to the natural conclusion that two $EOS$ with 
different effective masses at saturation,
 even if the compressibilities are the same, are expected to have 
different dynamical monopole response. 
In the $NLG-2$ vs. $NL3$ comparison discussed here we clearly see the 
interplay between compressibility and effective mass: we can get similar
monopole energies increasing the compressibility while decreasing the
effective mass, see Table \ref{table2}.

This appears to be a quite general feature, present also
in non-relativistic approaches, see the Fig.7 of the ref. \cite{Blaizot_2}
where the $RPA$ systematics of the Gogny forces is shown: the $^{208}Pb$
breathing mode energy is not much changing if at the same time we increase
the $NM$ compressibility and decrease the effective mass. A similar
trend has been suggested in a recent work on the nuclear compressibility
within the nonrelativistic frame \cite{colo}.

\subsubsection{Monopole frequency and symmetry energy}

It is well known that the equilibrium properties of nuclear matter
are changing with isospin asymmetry, in particular the saturation density
and the corresponding $EoS$ curvature , see \cite{BaranPR410},
 Sect.2 and refs. therein.
For the compressibility shift we have, after some algebra:
\begin{eqnarray}
\Delta K_{nm}(I) = 9\rho_0 \Big[ \rho_0 {\frac{d^2}{d\rho^2}}
 - 2 {\frac{d}{d\rho}} \Big] {\epsilon_{sym}} (\rho) \Big\vert
_{\rho=\rho_0}I^2 &&\nonumber\\
 = [K_{sym}-6L]I^2 ~<~0,
\label{decom}
\end{eqnarray}
with $I \equiv (\rho_n-\rho_p)/\rho$, asymmetry parameter. 
We note the interplay between slope, 
$L \equiv 3 \rho_0 {\epsilon_{sym}'} (\rho)$,
and curvature, $K_{sym} \equiv 9 \rho_0 {\epsilon_{sym}''} (\rho)$   
of the symmetry energy at saturation. The asymmetric matter becomes
softer since the shift is in general negative, due to the dominance of the
slope term $L$ \cite{curvature}, see \cite{BaranPR410} and the recent 
discussion 
in \cite{lwchen05}. Thus, in $^{208}Pb$, n-rich system, $I=0.21$, asymmetry 
can affect the the isoscalar monopole oscillations, as also noted in refs.
\cite{PiekaPRC66,AgraPRC68,colo} within relativistic and non-relativistic 
frames.

\begin{figure*}
\includegraphics[scale=0.6,angle=-90]{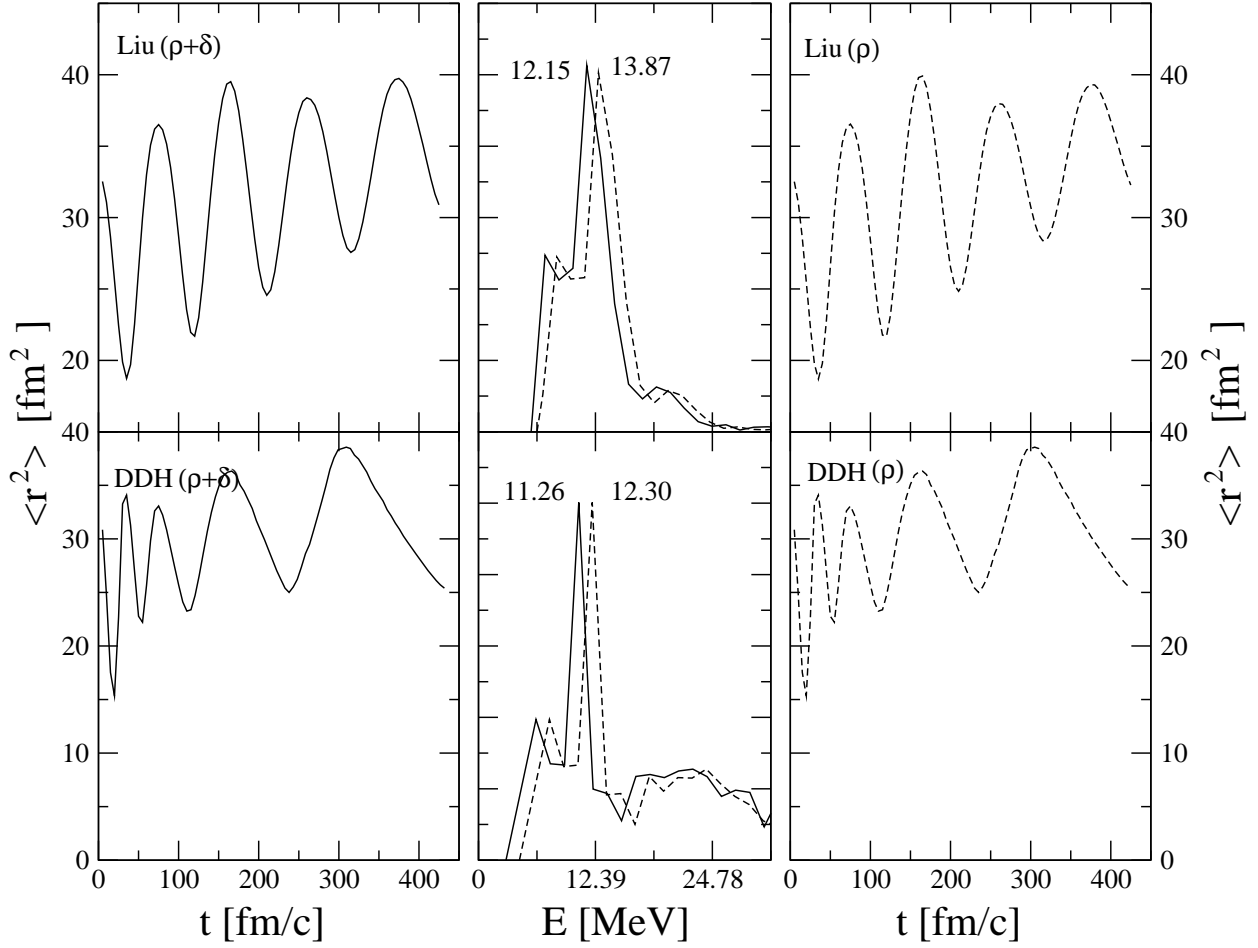}
\caption{\label{fig:2}  (Left and right pannels) Time-dependent isoscalar 
$\left\langle r^{2}\right\rangle$ monopole moment and (middle pannels) the 
corresponding 
Fourier power spectra for $^{208}$Pb. Transport calculations with the 
parameter sets are 
shown: (panels on the top)
$Liu-RMF$ with ($\rho$) dashed line (right)  and with ($\rho+\delta$) 
solid line (left). 
(panels on bottom) Density Dependent model ($DDH$) (lower panel) with ($\rho$) 
dashed line (right)  and with ($\rho+\delta$) solid line (left). 
The excitation energy of $ISGMR$ for each parameter set is explicitly 
indicated in the 
Fourier power spectra and for comparison the results with ($\rho+\delta$) and 
($\rho$) 
for both parameter set are shown together.}
\end{figure*}

In our $NL3$ vs. $NL2-G$ comparison the difference on the $a_4$ values,
see Table \ref{table2}, automatically implies a difference on the slope
parameter $L$, since in both models the potential symmetry energy is
coming from the effective $\rho$-meson coupling which leads to a linear
$\rho_B$-dependence. So the larger $a_4$ of $NL3$ means a larger slope at
$\rho_0$ and so a larger reduction of the compressibility in $^{208}Pb$,
 see Eq.(\ref{decom}). Our conclusion is that in $NL3$ both effects,
smaller effective mass and larger $a_4$, are almost compensating
the much larger $K_{nm}$ value, finally leading to a monopole frequency
in $^{208}Pb$ not much different from the $NL2-G$ one. 

We have continued the analysis of the symmetry contribution studying the 
effect of a scalar isovector channel with the inclusion of 
$\delta$ meson both in nonlinear and density dependent models. The results 
are shown in Fig.\ref{fig:2}.
The time history of isoscalar monopole moment and its Fourier spectrum are
 shown for the $Liu$ sets in the upper panel with isovector $\rho+\delta$ 
interaction
 ($NL\rho\delta$, solid line) and with $\rho$ interaction 
($NL\rho$, dashed line) and similarly with 
density dependent models (lower panel). For both interactions although 
the time
 evolution shows a very similar behaviour the power spectrum presents a net 
reduction in the energy centroid when one includes the $\delta$ meson.
This is a very nice indication of the dominance of the $''Slope''$ of the
symmetry energy on the compressibility shift. In fact when the 
$\delta$-channel is included we have a clear increase of the symmetry energy
slope $L$ around saturation, see the discussion after Eq.(6-19) of 
ref. \cite{BaranPR410}; e.g. it results about $20\%$ larger in the
$Liu$ parametrizations.

 Our discussion suggests that although the inclusion of $\delta$ meson 
does not 
produce important effects on the ground state of finite nuclei 
\cite{Gaitanos},
 it has an interesting influence on the collective excitations of 
charge asymmetric finite nuclei, in particular for $^{208}Pb$.
 This appears a good suggestion for new experiments aiming to a better
determination of the poorly known slope (and curvature) of the symmetry term
around saturation.

In general the $DDH$ monopole frequencies are systematically below
the corresponding $Liu$ ones. However this appears mostly a joint 
effect of different effective masses and symmetry energies, as already
discussed for the $NL3$ vs. $NL2-G$ comparison.
Therefore the density dependence of the coupling constants seems not largely
affecting the monopole response. In fact this can be expected
from the rather smooth behaviour around $\rho_0$, see Fig.1 of
ref. \cite{Gaitanos}. A more detailed study is presented in the Section V. 

Finally we like to note again that, although the $Liu$'s 
$NL\rho$ parametrization
 has a smaller $K_{nm}$ compared to $NL3$, it produces roughly the same 
monopole main frequency  
due to a larger effective mass joint to a smaller $a_4$. Of course
we cannot compare the $NL\rho\delta$ results since the $\delta$ channel
is absent in $NL3$.

\section{\label{sec:level4}ISOVECTOR DIPOLE OSCILLATIONS}

In the literature only Relativistic $RPA$ calculations have been performed 
so far to study the well known 
isovector giant dipole resonance $(IVGDR)$ \cite{Ma,Niksic}. It has been 
reported that for calculations of higher multipole modes, other than monopole,
the response of a nucleus is difficult to evaluate in the 
time-dependent $RMF$ \cite{Ma}. The reason for this is that, since rotational 
invariance is broken and the differential equations have to be explicitly 
solved at each time-step on a two dimensional mesh in coordinate space, it
 becomes difficult to keep the solutions stable for the very long times that
 are necessary for a good accuracy.  The problem is overcome in 
the present $RBUU$ simulation, since it is possible to study the time 
history of $IVGDR$ in a time dependent frame as already done in the $ISGMR$
case.

\begin{figure*}
\includegraphics[scale=0.6,angle=-90]{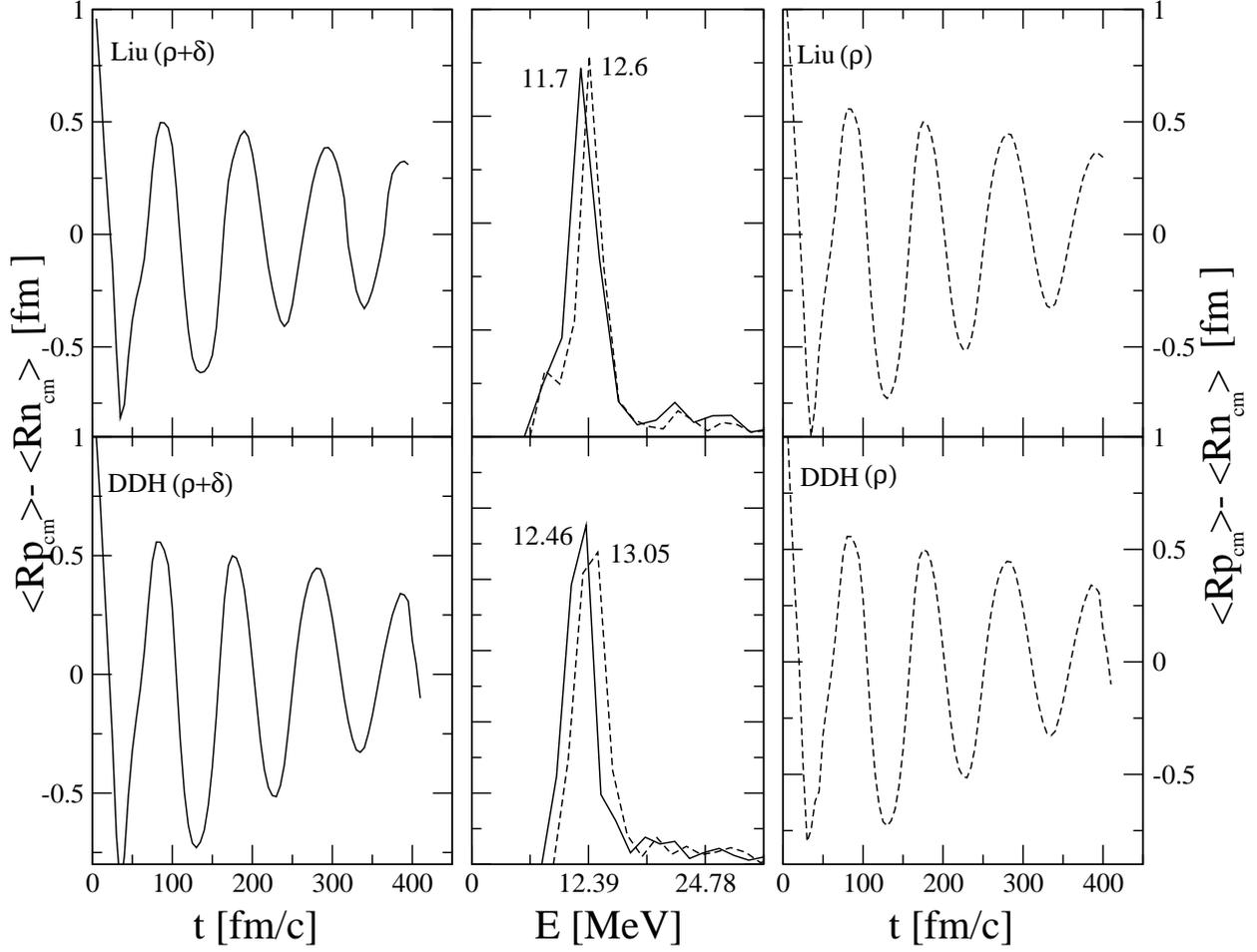}
\caption{\label{fig:3} Time-dependent isovector dipole moment (center 
of mass displacement of protons and neutrons 
$\left\langle Rp_{cm}\right\rangle -\left\langle Rn_{cm}\right\rangle $) 
and the corresponding Fourier power spectrum for $^{208}$Pb. The parameter set 
of the effective Lagrangian is $''Liu''$ (upper panel) with ($\rho$) dashed 
line (right)  and ($\rho+\delta$) solid line (left), and Density Dependent 
model $(DDH)$ (lower panel) with ($\rho$) dashed line (right)
  and ($\rho+\delta$) solid line (left). The excitation energy of $ISGMR$ 
for each parameter set is explicitly stated on the Fourier power spectrum and 
for comparison the results with ($\rho+\delta$) and ($\rho$) for both 
parameter sets are shown together.}
\end{figure*}

The experimental IVGDR energy in $^{208}$Pb is well known as 
13.5$\pm$0.2 MeV \cite{Berman}. Since in the isovector channel we 
mainly want to study the effect of the 
inclusion of 
isovector-scalar couplings, we only present results of the models that are 
parametrized with and without  $\delta$ meson. 
In Fig. \ref{fig:3} we present the $^{208}$Pb $IVGDR$ oscillations
 and the corresponding Fourier transforms within models that include 
isovector-scalar channel namely $Liu$ (upper panel) with $\rho+\delta$,
 solid line(left) and $\rho$, 
dashed line (right) and similarly for Density Dependent parameter sets 
(lower panel).
 The Fourier transforms show
a good single frequency dominance of the isovector mode. We observe that the 
$DDH$ models, with $a_{4}=33.4~MeV$, systematically give a larger resonance 
energy compare to $Liu-RMF$ sets, with 
$a_{4}=30.5$ MeV. Moreover in both models a clear reduction of centroid 
energy is observed when the $\delta$-meson is included.

It is a well known fact that the dynamics of $IVGDR$ is rather sensitive to 
the symmetry energy of the corresponding model which is acting as a kind of
restoring force parameter. 
The drawback in previous relativistic models is the one-to-one correspondence
between $a_4$ and $IVGDR$ energy \cite{Ma}, and so it is difficult to 
discriminate among the different interactions.
This is  not the case in our analysis.
The new important conclusion that can be drawn from
our results is that the $IVGMR$ dynamics is also sensitive to the more 
microscopic covariant structure of the symmetry term, i.e. to the interplay
of various isovector channels.
The physical interpretation of this 
result can be given in terms of the isovector response derived 
Ref.\cite{Greco_1}, where it is shown that the potential part of the symmetry 
energy explicitly appears in the dispersion relations with an correction term 
having a definite $f_{\rho}$, $f_{\delta}$ structure:
\begin{equation}\label{dipdis}
1+ \frac{6\,E^*_F}{k^2_F}\left[E_{sym}^{pot}-\frac{f_\rho}{2}
\frac{k^2_F}{E^{*2}_F}\left(1-f_\delta\,\frac{m^*}{E^{*2}_F}\,\rho_S\right)
\rho_B\right]\varphi(s)=0\,
\end{equation}

Note the similarity with the corresponding dispersion relation for 
isoscalar modes Eq.(\ref{mondis}), in particular the parallel role played by 
comressibility and symmetry energy in the two collective degrees of freedom.

Now {\it we can easily have interactions with the same
$a_4$ value at normal density but with rather different isovector dipole
 response}.
In fact when including the $\delta$ channel we have to increase the
$f_\rho$ coupling in order to have the same symmetry energy at saturation
($\delta$, scalar field, is attractive in the isospin degree of freedom,
see the discussion in Sect.6 of ref.\cite{BaranPR410}). 
Then the isovector dipole `` effective restoring force'' 
(coefficient of the 
Lindhard function in the Eq.(\ref{dipdis}) will be  reduced.

A similar effect has been pointed out in
a detailed non-relativistic $Skyrme-RPA$ study of the Giant Dipole
Resonance in heavy nuclei ($^{208}Pb$) using effective interactions
with various isovector terms \cite{ReinhardNPA649}. 
A separate sensitivity
of the average resonance frequencies on the symmetry energy $a_4$
and on its slope has been found. In a covariant scheme we can see from
Eq.(\ref{dipdis}) that such behaviour can be achieved only by using two 
isovector fields, at the lowest order.
This result shows more generally that a dynamical observable can be
more sensitive to the microscopic structure of the isovector interaction
than static properties.
For instance in a careful study of the neutron distributions,
 \cite{FurnstNPA706},
 it is clearly shown that these ``equilibrium'' observables are
almost equally correlated to value, slope and curvature of the symmetry
term.

Finally we see that even for the isovector dipole we cannot reveal specific 
contributions related to a density dependence of the coupling constants, in 
the sense that all the observed differences can be accounted for just
in terms of symmetry energies and of isovector channels. We must say
that we have not really performed an accurate analysis of this point, e.g.
comparing with $RMF$ models with exactly the same saturation properties.
Enlightening analitical results can be derived within a nuclear matter 
linear response theory, as shown in the following Section. 

\section{\label{sec:vdd}Nuclear Matter Response with Density Dependent 
Vertices}

In previous Sections in order to explore the relevant physical quantities  
affecting the energy centroid of the collective modes,
we have compared results from various $QHD$ models and parametrizations. 
Furthermore to have a guidance on the observed effects in finite nuclei,
we took advantage of the plain dispersion relations for collective modes 
in nuclear matter studied in Ref. \cite{Greco_1} in the contest of $QHD-NL$
models. However in the $DDH$ approach, where meson-nucleon couplings depend on 
the vector (baryon) density, the relation between the coupling functions 
and the 
compressibility can be expected to be modified together with the dispersion 
relations for isoscalar and isovector modes. 

By a comparison of results with the different parametrizations,
we have argued that the density dependence of
meson-nucleon couplings do not carry specific contributions at
least to a large extent. 
In order to corroborate such a statement we briefly discuss the pertinent 
modifications of the thermodynamic quantities together with the linear 
response theory in nuclear matter, showing that in $DDH$ models most of 
the effect is reabsorbed by
the compressibility and large corrections may be expected only for 
very strong density dependence of the couplings.

We remind that the $DDH$ model has all the meson-nucleon 
couplings dependent on the isoscalar vector (baryon) density, defined as
$\varrho=\sqrt{j_\mu j^\mu}$, with $j_\mu$ defined as in Section 
\ref{sec:level2}. Such a density dependence leads to rearrangement 
contributions
that affect the vector self-energy adding a term that at mean field level 
reads as:
\begin{equation}
\label{rear}
\Sigma_\mu^R= 
\left(\frac{\partial f_\omega}{\partial \varrho}\varrho^2 -
\frac{\partial f_\sigma}{\partial \varrho}{\rho_s^2} \pm 
\frac{\partial f_\rho}{\partial \varrho}j_{3\mu}j_3^{\mu}\mp 
\frac{\partial f_\delta}{\partial \varrho}\rho_{s3}^2\right)\frac{j_\mu}
{2 \varrho}.
\end{equation}
The effect of rearrangement terms on the pressure can be find in refs.
\cite{FuchsPRC52,Typel}. More relevant for our discussion
is how they enter the compressibility:
\begin{equation}
\label{compr-vdd}
K=3\frac{k_F^2}{E^*_F}+9\,\left(f_\omega\rho_B-\,f_\sigma\frac{m^*}{E^*}
\frac{d\rho_s}{d\rho_B}\rho_B \right)
+9\,\left(\frac{df_\omega}{d\rho_B}\rho_B^2-
\frac{df_\sigma}{d\rho_B}\frac{m^*}{E^*}\rho_s \,\rho_B \right)+
9\,\frac{d\Sigma_0^R}{d\rho_B}\rho_B
\end{equation}
with 
\begin{equation}
\frac{d\Sigma_0^R}{d\rho_B}=\left(\frac{df_\omega}{d\rho_B}\rho_B-
\frac{df_\sigma}{d\rho_B}\frac{d\rho_s}{d\rho_B}\rho_s\right)
+\frac{1}{2}\left(\frac{d^2f_\omega}{d\rho_B^2}\rho_B^2-
\frac{d^2f_\sigma}{d\rho_B^2}\rho_s^2 \right)
\end{equation}
we can see that Eq.(\ref{compr-vdd}) reduces to the standard formula 
of $QHD$ with costant couplings when the coupling are density indipendent, 
see Ref. \cite{BaranPR410}.

The linear response in nuclear matter can be derived along the same lines
of Ref.\cite{Greco_1}, but one as to redefine the vector self-energy
$\Sigma_\mu$, defined in Eq.(\ref{self-energy}), adding the 
rearrangement term $\Sigma_\mu^R$ and consequently the effective 
kinetic momenta, defined in  
Eq.(\ref{peff}). Once also the equilibrium Wigner
matrix is consistently redefined, on the same fashion of Ref.\cite{Greco_1}
a dispersion relation for isoscalar modes in symmetric nuclear matter
can be written as:

\begin{equation}\label{mondis-VDD}
1+ \frac{E^*_F}{3\,k_F^2}\left[K_{nm}^{pot}-9\,f_{\omega}
\frac{k^2_F}{E^{*2}_F}\left(1-f_\sigma\,\frac{m^*}{E^{*2}_F}\,\rho_s
 + \frac{\rho_s^2}{E^{*}_F}\,\frac{df_\sigma}{d\rho_B} \right)
\rho_B  \right]\varphi(s)=0\,.
\end{equation}
if terms associated with $\frac{\partial \rho_s}{\partial m^*}$
are discarded, or in other
words, if the safe approximation $\frac{d\rho_s}{d\rho_B}=\frac{m^*}{E^*}$
is taken.
We can see that of the derivative of the coupling function are 
reabsorbed in the $K_{nm}^{pot}$. The difference between Eq.(\ref{mondis})
and Eq.(\ref{mondis-VDD}) is the last term. With some algebra, within the same
approximation scheme,  the contribution
of density dependent terms can be reduced to a variation of the 
$\sigma$-coupling, from $f_\sigma$ to 
$f_\sigma-\frac{df_\sigma}{d\rho_B}\,\rho_B$, 
 in the correction term of Eq.(\ref{mondis}). Due to the other 
quenching factors
and to the smoothness of the $f_\sigma(\rho_B)$ function around $\rho_0$,
 we can expect an overall variation of a few percent in the dispersion 
relation.

We note however 
that now the $f_\omega (\rho_0)$ is not exactly the same of 
$QHD-NL$ (with scalar non-linear terms), because the rearrangement terms affect
also the relation for the binding energy, that in $DDH$ models is given by:
\begin{equation}
E/A(\rho_0)=E^*_F + f_\omega\rho_0 + \Sigma_0^R - m_N
\end{equation}
From Eq.(\ref{rear}) we see that
 $\Sigma_0^R$ is given by the difference in the density slope
 between the scalar and vector field. Therefore it
may not be discarded in general, but is usually subdominant due to the similar
density dependence of the scalar and vector coupling functions
around saturation density, see Fig.1 of ref.\cite{Gaitanos}.

The expression for the symmetry energy 
is not modified in $DDH$ models respect to $QHD-NL$ \cite{Liu_1}, because the 
couplings 
depend only on the isoscalar vector density. Also the dispersion relation for 
the isovector
mode in symmetric nuclear matter is not affected by the density dependence of
 the couplings and the relation written in Eq.(\ref{dipdis}) is still  valid.
However in asymmetric nuclear matter new terms appear, that are proportional 
to the asymmetry $I$ and to the derivative of the $\rho-$like and 
$\delta-$like couplings. 
Therefore in $DDH$ models for exotic nuclei there is the possibility to have
 a modified relation between symmetry energy and dipole excitation, however 
an exhaustive study of such effects goes beyond the scope of this paper.

\section{\label{sec:level5}CONCLUSIONS}
Aim of this work has been to study effects on collective excitation
 properties of asymmetric spherical nuclei of the detailed structure of the
nuclear $EoS$ in the isoscalar and isovector sector.
Particular attention has been put on the dynamical implications of the
inclusion of a scalar isovector contribution due to an effective $\delta$
meson coupling.

We have described the dynamics of the isoscalar monopole and isovector dipole 
oscillations in a relativistic transport approach based on a nucleon-meson 
effective field interaction. We have evaluated the time-evolution of the 
oscillations  
obtaining then the corresponding  excitation energies from the power 
spectra of the modes. The applications are to the n-rich nucleus $^{208}Pb$,
using a suitable choice of different effective Lagrangians.

For the monopole mode we show an interesting $m^*$, effective mass, dependence
of the centroid energy. This effect, joint to a symmetry energy contribution,
 can account for the claimed ambiguity on the difference in compressibility
 moduli, that produce the experimental energy of $ISGMR$ in $^{208}Pb$ , 
between relativistic and nonrelativistic microscopic models.
From the influence of the $\delta$ meson, just a systematic reduction of the
peak energy, we have inferred that the symmetry energy effect on the isoscalar 
monopole is mainly due to variations of the $''Slope''$-$L$ around saturation.
This is an important result, since it could open the possibility of a direct
access to this poorly known parameter from monopole data changing the
neutron number in a fixed isotope. Moreover we will be able to trace back
the covariant structure of the effective interaction in the isovector channel,
 of relevance also for relativistic heavy ion collisions \cite{Gaitanos}.

The isovector dipole response is directly linked to the isospin dependent part
of the nuclear $EoS$. The new result shown here is that the $IVGDR$ energy
is decreasing when the $\delta$ coupling is included, keeping fixed the
symmetry energy at saturation. The effect can be easily understood from
the form of isovector dispersion relations in nuclear matter in a 
relativistic
linear response approach. We can then have different $IVGDR$ energies for
effective interactions that show the same $a_4$ value, but with a different
covariant field structure in the isovector channel. 
We note the similarity to the isoscalar case, just exchanging the roles
of compressibility and isoscalar couplings with symmetry energy and
isovector couplings.
 
Finally we do not see large effects from the Density Dependence of the 
effective meson couplings. This is mainly due to the smooth behavior
around $\rho_0$. Moreover we have shown that some compensation 
is also coming from dynamical contributions of the
rearrangement terms.

\begin{acknowledgments}
S.Y would like to thank the staff of Laboratori Nazionali del Sud 
(LNS), INFN, 
in Catania for the hospitality during his visit to LNS. This work has been 
supported in part by Mersin University under the contact No. BAB FEF FB (SY)
 2004-2.
\end{acknowledgments}

%
%
%


\begin{thebibliography}{99}

\bibitem{Liu_1}B.Liu, V.Greco, V.Baran, M.Colonna, M.Di Toro, 
Phys.Rev.  {\bf C65}, 045201 (2002).
\bibitem{BaranPR410} V.Baran, M.Colonna, V.Greco, M.Di Toro,
 Phys.Rep. {\bf 410}, 335-466 (2005).
\bibitem{Greco_1}V.Greco, M.Colonna, M.Di Toro, F.Matera, Phys.Rev. 
{\bf C67}, 045201 (2003).
\bibitem{Ring} P.Ring, D.Vretenar and B.Podobnik, Nucl.Phys. {\bf A598}, 
 107 (1996).
 
\bibitem{Lalazissis} G.A.Lalazissis, J.Konig and P.Ring, Phys.Rev. 
{\bf C55}, 540 (1997).
\bibitem{FuchsPRC52} C.Fuchs, H.Lenske, H.H.Wolter, Phys.Rev. {\bf C52},
 3043 (1995).
\bibitem{Typel} S.Typel, H.H.Wolter, Nucl.Phys. {\bf A656}, 331 (1999).
\bibitem{Stoitsov_1} M.V.Stoitsov, P.Ring, and M.M.Sharma,
 Phys.Rev. {\bf C50}, 1445 (1994).
\bibitem{Vretenar_1} D.Vretenar, N.Paar, P.Ring, and T.Niksic,
 Phys.Rev. {\bf C65}, 021301 (2002).
\bibitem{Vretenar_2} D.Vretenar T.Niksic, and P.Ring, Phys.Rev. {\bf C68}, 
024310 (2003).

\bibitem{Blaizot_1} J.P.Blaizot, Phys.Rep. {\bf 64}, 171 (1980).
\bibitem{Blaizot_2} J.P.Blaizot, J.F.Berger, J.Decharge and M.Girod, 
Nucl.Phys.  {\bf A591}, 435 (1995).
\bibitem{Vretenar_4} D.Vretenar, G.A.Lalazissis, R.Behnsch, W.Poschl and 
P.Ring, Nucl.Phys.  {\bf A621}, 853 (1997).
\bibitem{Ma} Z.Y.Ma, N.Van Giai, A.Wandelt, D.Vretenar and P.Ring,
 Nucl.Phys.  {\bf A686}, 173 (2001).

\bibitem{PiekaPRC66} J.Piekarewicz, Phys.Rev. {\bf C66}, 034305 (2002).

\bibitem{AgraPRC68} B.K.Agrawal, S.Shlomo, V.Kim Au, Phys.Rev. {\bf C68}, 
 031304(R) (2003).

\bibitem{colo} G.Col\'o, N.Van Giai, J.Meyer, K.Bennaceur and P.Bonche
 Phys.Rev. {\bf C70}, 024307 (2004).

\bibitem{qhd} B.D.Serot,J.D.Walecka, Adv.Nucl.Phys. {\bf 16}, 1 (1986).

\bibitem{Botermans_1} W.Botermans and R.Malfliet, Phys.Rep. {\bf 198}, 
 115 (1990).
\bibitem{Blaettel} B.Blaettel, V.Koch, U.Mosel, Rep.Prog.Phys. {\bf 56}, 
 1 (1993).
\bibitem{Danielewicz} P.Danielewicz, Ann.Phys. {\bf 152}, 239 (1984).

\bibitem{Gaitanos} T.Gaitanos, M.Di Toro, S.Typel, V.Baran,  C.Fuchs, V.Greco,
 H.H.Wolter, Nucl.Phys.  {\bf A732}, 24 (2004).

\bibitem{Maruyama}T.Maruyama, W.Cassing, U.Mosel, S.Teis, K.Weber, Nucl.Phys.
 {\bf A573}, 653 (1994).

\bibitem{Fuchs_1} C.Fuchs, H.H.Wolter, Nucl.Phys.  {\bf A589}, 
 732 (1995).

\bibitem{Stocker_1} H.St\"{o}cker, W.Greiner, Phys.Rep. {\bf 8}, 
 137-277 (1986).

\bibitem{Shlomo} S.Shlomo and D.H.Youngblood, Phys.Rev.  {\bf C47}, 52 
(1993).
\bibitem{Vretenar_7} D.Vretenar, N.Paar, P.Ring, G.A.Lalazissis,
 Nucl.Phys. {\bf A692},  496 (2001).
\bibitem{Niksic} T.Niksic, D.Vretenar, P.Ring, Phys Rev. {\bf C66}, 064302 
(2002).

\bibitem{Podobnik} B.Podobnik, D.Vretenar and P.Ring, Z.Phys. {\bf A354},
 375 (1996).

\bibitem{CT}
 A.Smerzi, A.Bonasera, M.Di Toro, Phys.Rev. {\bf C44}, 1713 (1991).

\bibitem{curvature} In fact the symmetry energy curvature is also negative
for most effective interactions (we have also a negative contribution from 
the kinetic part). It can be slightly positive only for a highly increase,
 of parabolic type, of the potential contribution around the saturation
point, ref.\cite{BaranPR410}.

\bibitem{lwchen05} L.-W.Chen, C.M.Ko, B.-A.Li, Phys.Rev.Lett. {\bf 94},
 032701 (2005).

\bibitem{Berman} B.L.Berman and S.C.Fultz, Rev.Mod.Phys. {\bf 47}, 713 (1975).

\bibitem{ReinhardNPA649} P.G.Reinhard, Nucl.Phys. {\bf A649}, 305c (1999).

\bibitem{FurnstNPA706} R.J.Furnstahl, Nucl.Phys. {\bf A702}, 85 (2002).


\end{thebibliography}
\end{document}